# Spectrally resolved white light interferometer for measuring dispersion in the visible and near infrared range

Yago Arosa, Elena López-Lago, Raúl de la Fuente

Grupo de Nanomateriais, Fotónica e Materia Branda, Departamentos de Física Aplicada e de Física de Partículas, Universidade de Santiago de Compostela, E-15782, Santiago de Compostela, Spain

ABSTRACT

We design a spectrally resolved interferometer to measure the refractive index of transparent samples over a broad spectral range of 400–1550 nm. The measuring device consists of a Michelson interferometer whose output is analysed by means of three fibre spectrometers: a homemade prism spectrometer which obtains the interferogram generated by the sample in the 400–1050 nm range, a homemade transmission grating spectrometer that measures the interferogram in the 950–1550 nm near infrared range, and a commercial Czerny-Turner spectrometer used to make high-precision measurements of the displacement between the Michelson mirrors. The whole system is illuminated by a white-light source with an emission spectrum similar to that of a black body. We test the instrument on solid and liquid samples achieving accuracy of up to $10^{-4}$ in the refractive index after fitting it with the Cauchy formula.

## 1. Introduction

In optics, dispersion refers to the dependence of the optical properties of a material or device upon wavelength. Material dispersion, or chromatic dispersion, which describes the wavelength dependence of the refractive index, concerns any phenomena related to refraction and therefore affects the efficiency of many optical systems operating in different spectral intervals. One important example is optical fibre communications where the data bit rate would be strongly limited if the broadening of the signal pulse due to dispersion during propagation was not minimized or compensated [1]. Dispersion also affects the operation of nonlinear optical devices: it is responsible for the temporal walk off, its characterization is essential to determine the phase-matching condition [2], and it can minimize the efficiency of four-wave mixing processes [3], to list some examples. In addition, it leads to chromatic aberration that affects optical imaging systems and especially optical microscopy [4]. The photonic crystal technology is affected by dispersion as well [5]; in particular, the guiding properties of photonic crystal fibres filled with liquids are very sensitive to the chromatic dispersion of the liquid [6]. Different endeavours in the laser technology [7], such as, pulse compression or propagation of increasingly shorter pulses, require precise knowledge of the behaviour of chromatic dispersion in certain wavelength ranges in the VIS-NIR region.

Therefore, accurate knowledge of dispersion is crucial when operating at different frequencies in the VIS-NIR spectral region in many fields of science and industry, including optical design, optical imaging, optical communication, laser physics, low-coherence metrologies, and ultrafast optics. Additionally, dispersion can be used for sensing [8]; to obtain information about various physical properties and chemical composition, for example, impurities content or environmental conditions; it is also essential for the development of theoretical and numerical physical models.

From the pioneering work of Sainz and co-workers [9–11], the analysis of interference of incoherent light in the spectral domain (spectrally resolved white-light interferometry, SRWLI) has been shown to be a powerful tool to measure material dispersion over a broad spectral range [12–21]. The characterization of materials requires covering the widest possible spectral range while preserving high resolution, a challenge that must be overcome in a proper way. For example, Hlubina [21] used a low-resolution spectrometer to obtain the group and differential group indices over the 500–900 nm spectral range by sequentially evaluating the stationary phase point position as a function of the interferometer path length difference in air; the achieved accuracy for the differential group index was ~ $7 \times 10^{-5}$. Delbarre et al. [15] processed three interferograms including a stationary phase point to get the group index and estimate the refractive index from 540 to 660 nm; the relative accuracy was ~ $9 \times 10^{-4}$ for the group index and $5 \times 10^{-4}$ for the refractive index. Reolon et al. [17] used a broadband supercontinuum source with high degree of spatial coherence to increase the fringe visibility and measured the group index from 530 to 800 nm in a single acquisition, attaining an accuracy of $3 \times 10^{-4}$. Recently [20], we measured the refractive index of solid samples in a single acquisition in the spectral range of 400–1000 nm. The accuracy of the results was ~ $10^{-4}$ for the group index and below $10^{-4}$ for the refractive index. We also applied SRWLI to measure the refractive index of 14 imidazolium-based ionic liquids [22] in the same spectral range with accuracy better than $2 \times 10^{-4}$.

In this paper, we describe the measurement of the dispersion of solid and liquid samples in a very broad range of 400–1550 nm. To achieve this goal, we incorporate a second spectrometer into our apparatus to obtain data for long wavelengths. In Section 2, we review the principles of SRWLI, highlighting the main issues arising when applying this technique; in Section 3, we describe our device; in Section 4, we detail the different steps of the experimental procedure; in Section 5, we present and discuss the results obtained with different samples; finally, in Section 6, we present our conclusions.

## 2. Theoretical foundation

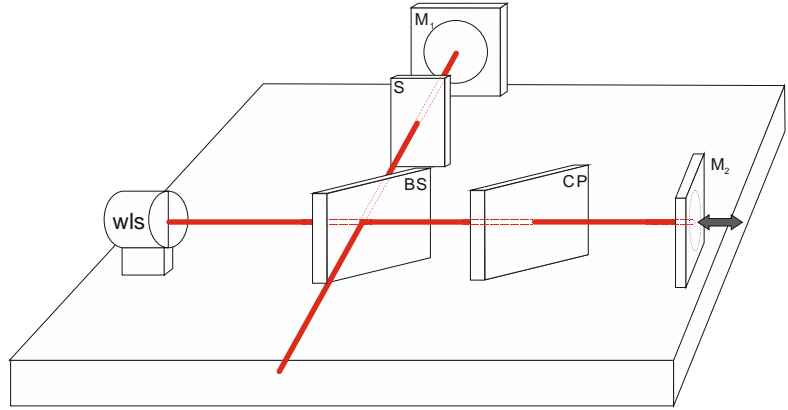

Fig. 1 Typical Michelson interferometer with the sample, S, and a fixed mirror, $M_1$, in one arm and a moving mirror, $M_2$, in the other arm. WLS, white light source; BS, beam splitter; CP, compensating plate.

In this section, we consider a Michelson interferometer (see Fig. 1) with the sample to be measured in one arm (the sample arm) and a moving mirror in the other arm (the reference arm), which compensates the dispersion generated by the sample. At the output of the interferometer, we obtain the irradiance resulting from the interference of the beams travelling in its two arms. Considering light of wavelength $\lambda$, a sample of thickness $d$, and a displacement in air between the interference mirrors $l$, we get:

$$I(\lambda) = I_0(\lambda)\left[1 + V(\lambda)\cos\varphi(\lambda)\right], \tag{1a}$$

$$\varphi(\lambda) = 4\pi\left[d(n - n_{air}) - n_{air}l\right]/\lambda, \tag{1b}$$

where $I_0(\lambda)$ is the background spectral irradiance, $V(\lambda)$ is the fringe visibility, $\varphi(\sigma)$ is the phase difference between the beams of the interferometer, and $n_{air}$ and $n$ are the refractive indices of air and the sample, respectively, calculated at wavelength $\lambda$. When the interferometer is illuminated by a broadband source, some spectral components will experience maximum transmission, while others will be reflected by the beam splitter toward the source, and the remaining components will be partially transmitted and partially reflected. Therefore, the spectral irradiance exhibits an oscillating pattern with a varying period depending on the refractive index dispersion in the sample and air, as shown in Fig. 2. This rapidly varying irradiance can be easily evaluated by a spectrometer with sufficient resolution.

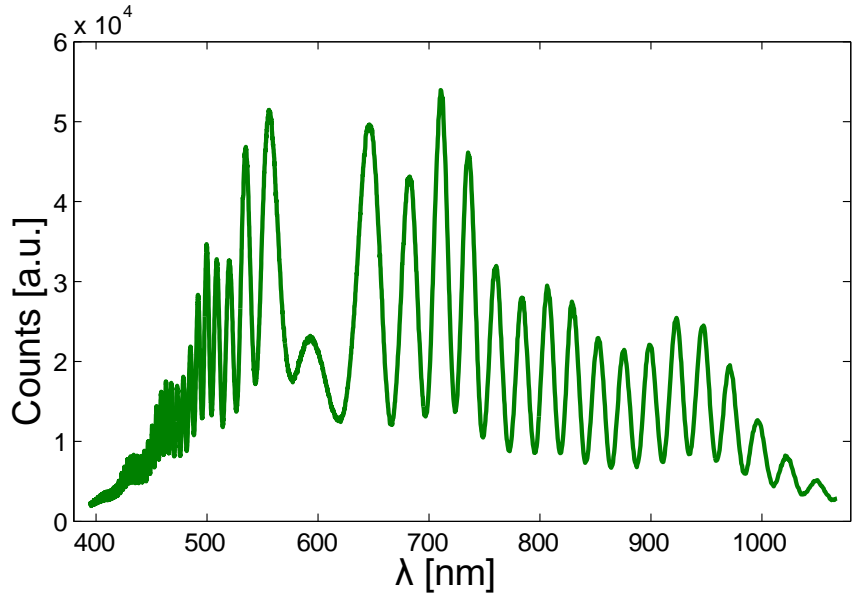

Fig. 2 Typical spectrogram obtained with SRWLI, showing a stationary phase point at a wavelength of ∼ 590 nm.

2.1. Phase evaluation

To obtain information about the sample dispersion, the phase must be extracted from Eq. (1a). There are a variety of methods to accomplish this task, such as, the Hilbert, Fourier, and wavelet transforms, phase shifting methods, or phase calculation from minimum and maximum detection [24–26]. However, regardless of the method which is applied, and because the inverse trigonometric function is multivalued, we cannot extract the exact phase in Eq. (1) and only have access to its principal value defined in the interval [−π, π]. Thus, we must apply an unwrapping procedure to eliminate discontinuities; however, even in this case, the extracted phase, $\varphi_u$, differs from the correct one by a multiple of $2\pi$, which can be written as:

$$\varphi_u(\lambda) = 4\pi \left[ d(n - n_{air}) - n_{air} l \right] / \lambda - 2k\pi, \; k \in \mathbb{Z}. \tag{2}$$

Hence, to solve the equation for the refractive index, $n$, we must first calculate the value of $k$. Previously, there have been some attempts to overcome the phase ambiguity and obtain the value of $k$ [11,14,15]. However, we believe that any method has an inherent error, and the correct approach is to calculate the refractive index separately at a particular wavelength, $\lambda_0$. Given the value of the refractive index at this wavelength, $n_0$, $k$ is calculated using Eq. (2) as:

$$k = \text{nint}\left\{ 2\left[ d(n_0 - n_{0air}) - n_{0air} l \right] / \lambda_0 - \varphi_u(\lambda_0)/2\pi \right\}, \tag{3}$$

where nint is the nearest integer function. By inserting $k$ into Eq. (2), the value of the refractive index can be found as a function of wavelength in the spectral range of interest. Of course, the sample thickness and the displacement between the interferometer mirrors must be determined first.

To obtain a reliable phase, the spectrometer must have sufficient resolution for resolving the rapidly oscillating spectral irradiance. This can be easily satisfied by choosing the value of *l* such that the spectrogram includes the stationary phase point. This point corresponds to the wavelength $\lambda_{eq}$, called the equalization wavelength, satisfying:

$$l = d \left[ \frac{n_g(\lambda_{eq})}{n_g(\lambda_{eq})|_{air}} - 1 \right], \quad (4)$$

where $n_g$ stands for the group index. Near the equalization wavelength, the oscillating spectral pattern varies slowly, and the oscillation period increases as we move away from that wavelength. Thus, as the mirror displacement, *l*, changes accordingly, the visibility in Eq. (1a) decreases and goes to zero for smaller or larger displacements. We note that we obtain the background irradiance at these points.

Unfortunately, many methods used to extract the phase in Eq. (1a) do not work well near the stationary phase point. Here, we use the procedure detailed in [20,25]. First, we separate the measured background irradiance in Eq. (1a); second, we obtain the upper and lower envelopes by interpolation of maxima and minima, respectively. In this step, the stationary phase point must be omitted. The visibility function is determined by subtraction of the two envelopes, it is removed from the equation and the cosine is extracted. Finally, the phase principal value is obtained by applying the arc cosine function.

The refractive index is obtained after applying an unwrapping method and calculating the value of *k* as described above. Once the refractive index is retrieved, the group index is estimated by differentiation.

## 3. The device

We design an SRWLI system which covers the spectral range of 400–1550 nm. In Fig. 3, we show the experimental configuration. Our SRWLI device is composed of five parts: a white light source, an interferometer, and three spectrometers. The white light source (WLS) is a halogen lamp with over 150 W of electric power, which can produce a black-body-like emission spectrum. The light beam is collimated by lenses L1 and L2 before entering the Michelson interferometer. The interferometer incorporates a mobile stage to place the sample and a heat exchanger (HE) that allows us to manage the temperature of the sample with a precision of up to 0.1 °C. When liquid samples are measured, they are placed in a quartz cell, and a compensating quartz plate (Q) is introduced into the interferometer reference arm. The output signal of the interferometer is analysed by two homemade spectrometers, a prism spectrometer (SWS) and a grating one (LWS), to retrieve the phase in the spectral region of interest, and a commercial Czerny–Turner spectrometer to measure *l*. It is essential to employ two different spectrometers to measure dispersion, because it varies distinctively in different spectral regions. Typically, the refractive index dispersion is higher for frequencies in the visible region than in the infrared region. As a consequence, the measurement in the VIS region requires a camera with a higher resolution than in the IR in order to resolve the rapidly varying interference fringes. For the same reason, we use a prism spectrometer to measure in the range of shorter wavelengths (up to 1 µm) and a grating spectrometer for longer wavelengths.

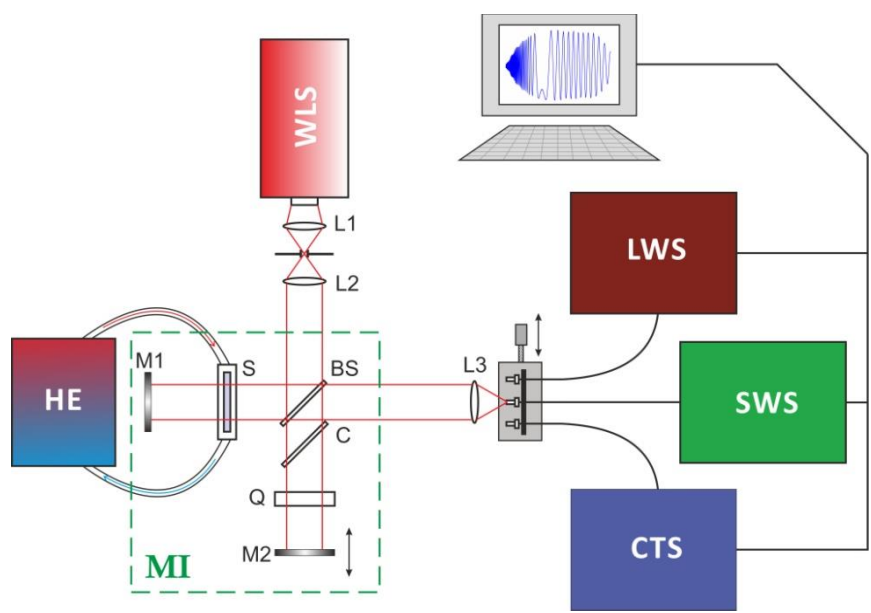

Fig. 3 Schematic of the optical set-up. WLS, white light source; MI, Michelson interferometer; SWS, short-wavelength spectrometer; LWS, long-wavelength spectrometer; CTS, Czerny–Turner spectrometer; HE, heat exchanger; L1 and L2, collimating lenses; L3, focusing lens; M1 and M2, Michelson mirrors; BS, beam splitter; C, beam splitter compensating plate; S, sample; Q, sample compensating plate.

The interferometer is connected to the spectrometers with optical fibres placed on a translation platform. The light is coupled into the fibres by focusing with an optical lens. It is not necessary to acquire data with the SWS and LWS spectrometers simultaneously since we average tens of interferograms, and the average remains constant in time.

3.1. Short-wavelength spectrometer

The first spectrometer covers the range of 400–1050 nm. It is a prism-based dispersion instrument that allows resolving fast irradiance oscillations at short wavelengths. We use an F2 dispersive prism because of its high optical quality, high refractive index, and low Abbe number and a high-resolution linear camera with 3648 pixels and 29.2 × 0.2 mm² area, as a detector. The focal length of the instrument is 315 mm.

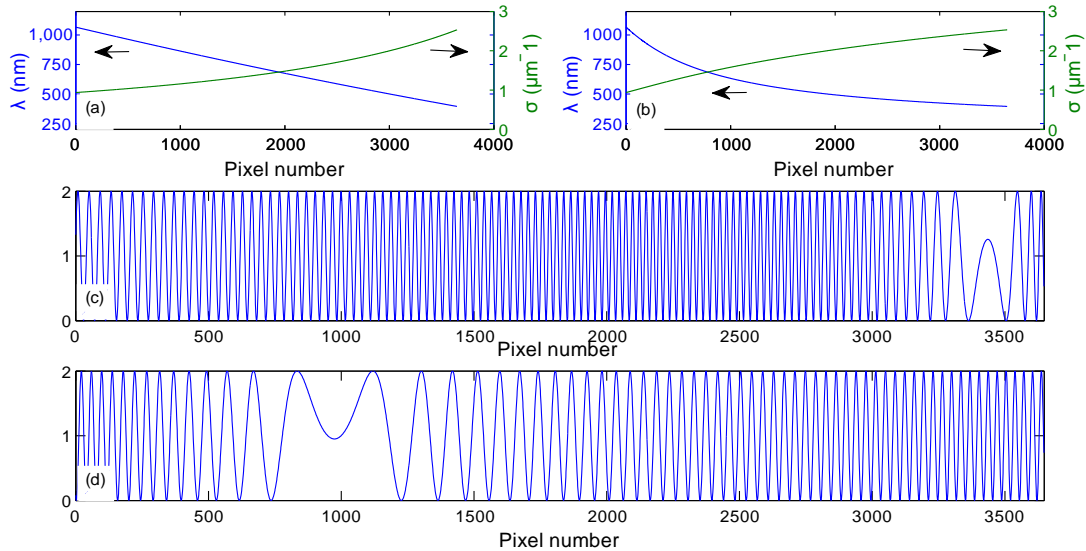

Fig. 4 Calibration of a typical grating spectrometer (a) and our homemade prism spectrometer (b). Plots (c) and (d) correspond to simulations of ideal interferograms, with constant background irradiance and maximum visibility, for the grating and prism spectrometers, respectively. In the simulations, it is assumed that there is a 1.5 mm thick fused silica plate in one of the arms of the Michelson interferometer. In each case, the difference in length between the interferometer arms is chosen to optimize the resolution of fringe detection.

Typically, commercial grating spectrometers used in SRWLI experiments do not resolve fringes in such wide spectral windows. Therefore, we assembled a prism spectrometer with a high-resolution camera to increase the visibility function. The difference between grating and prism spectrometers is the way of separating the spectra. In the former, the dispersion is nearly linear in wavelength, while in the latter, it is highly nonlinear in wavelength but more linear in wave number, $\sigma = 1/\lambda$, in air (see Fig. 4). In Figs. 4(c) and 4(d), we show the result of simulation of two ideal white-light spectral interferograms $[I(\lambda)=1+\cos\varphi(\lambda)]$ corresponding to the dispersion of a fused silica sample as they are seen in a typical grating spectrometer and in our homemade prism spectrometer. In the normal dispersion regime, the refractive index, and therefore the phase, varies rapidly at shorter wavelengths, meaning that for the grating spectrometer we must put the stationary phase point in the short-wavelength range, which corresponds to the right side of the sensor. However, since the phase varies quickly as we move away from the stationary phase point, the fringes cannot be resolved at the middle of the sensor, where dispersion is sufficiently large. On the other hand, the behaviour of the phase is closer to the sample dispersion in the prism spectrometer, and the blue part of the spectrum corresponds to a large part of the sensor. This allows resolving fringes even if they are far away from the stationary phase point and pushing it to low frequencies at the left-hand side of the camera. In consequence, the prism spectrometer resolves the fringe pattern much better, as seen in Fig. 4. Furthermore, prism spectrometers provide greater dispersion power for high frequencies in the visible region, where the material dispersion is typically higher, and consequently the refractive index variation is larger and lower dispersion for the IR region, where the sample refractive index varies slowly. Fig. 5 shows the spectral bandpass of the spectrometer for each pixel indicating higher dispersion at blue wavelengths.

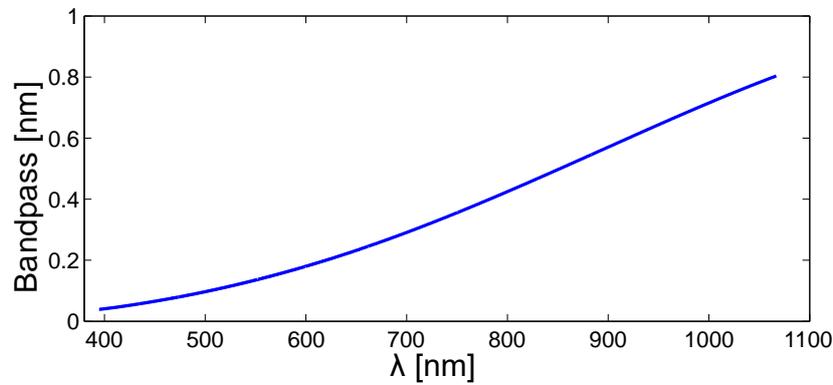

Fig. 5 Bandpass of the short-wavelength spectrometer as a function of wavelength.

Calibration of the spectrometer is performed with a Hg–Ar lamp which has many lines in the aforementioned spectral range. This lamp provides the relation between the pixel number and wavelength for a discrete number of pixels. The complete calibration for every pixel is achieved by applying SRWLI [26]. The resulting calibration is shown in Fig. 4(b).

3.2. Long-wavelength spectrometer

The second spectrometer covers the range of 950–1550 nm. We use a grating spectrometer in this case since materials typically show low dispersion in the NIR. This spectrometer uses a diffraction grating with 300 lines/mm and an infrared array camera with 320 × 256 pixels, 30 μm wide; the spectrometer has a focal length of 50 mm. Even though reflective gratings show better optical quality owing to reduced chromatic aberration, in our case, the spectrometer employs a transmission grating. This enables us to easily accommodate all the optical elements in such a short focal length instrument. Fig. 6 shows the calibration, nearly linear in wavelength, and the spectral bandpass of this spectrometer.

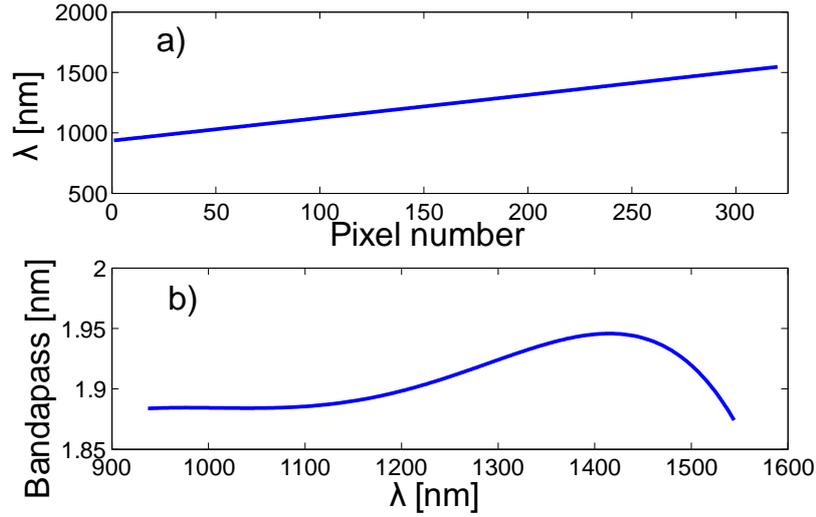

Fig. 6 Calibration of the long-wavelength spectrometer as a function of pixel number (a) and bandpass as a function of wavelength (b).

3.3. Czerny–Turner spectrometer

The third spectrometer is of the Czerny–Turner type, with 0.5 m focal length, a 1200 lines/mm reflective grating, and a CCD detector with 1024 × 128 pixels, 25 µm wide. It is used to measure the displacement between the mirrors in the Michelson interferometer, also by using SRWLI [27]. In our measurements, we displace the moving mirror in the reference arm of the interferometer to push the stationary phase point to the red band of the first spectrometer (~600 nm). Considering Eq. (4), for the group index ratio of 1.5 at the equalization wavelength and the sample thickness of 1 mm, we obtain $l$ = 0.5 mm. Without the sample, the phase in Eq. (1b) changes to $\varphi(\lambda) = 4\pi n_{air} l / \lambda$; if we neglect dispersion in air, the distance between successive maxima is $\Delta\lambda = \lambda^2 / (2 n_{air} l)$. For the spectral range of 0.4–1.55 µm, we find $\Delta\lambda$ = 0.16–2.4 nm, which is too small to be resolved with the first two spectrometers described above, meaning that they cannot be used to measure $l$ (although they could be used for sample thicknesses of several tenths of a millimetre). The Czerny–Turner spectrometer has a resolution of ~ 0.02 nm, sufficient for resolving the fringe spacing corresponding to the displacement between mirrors of several millimetres. For the measurement, we used a spectral window ~ 35 nm wide and centred at a wavelength of 800 nm. Previously, we performed our own spectrometer calibration. To obtain $l$, we approximate $n_{air}$ = 1. Thus, we can fit the phase to a linear function of wavenumber and determine $l$ from the slope.

## 4. Experimental procedure

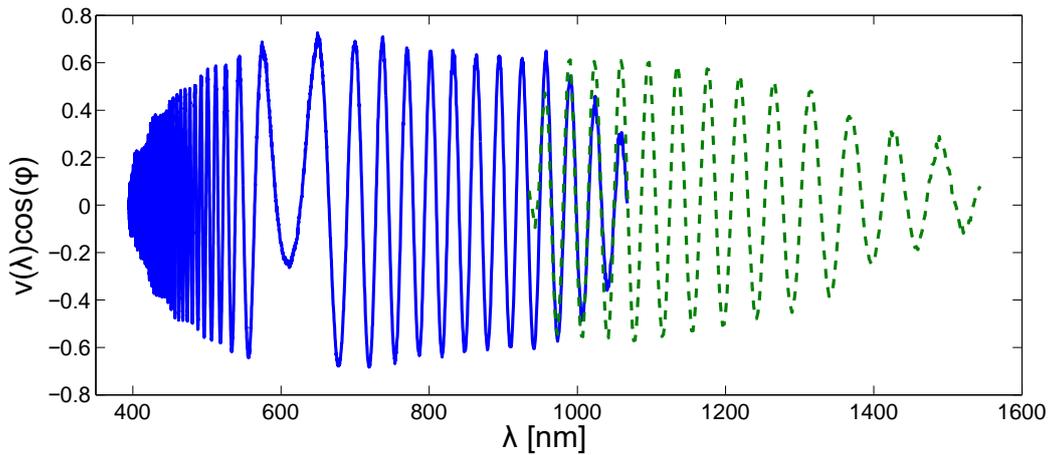

Fig. 7 Interference patterns from the two spectrometers as functions of wavelength: prism spectrometer (solid line) and grating spectrometer (dashed line).

The quality of the instrument is checked by measuring dispersion of samples with known refractive indices. In this work, we consider two solid (fused silica and BK7) and two liquid (deionised water and ethanol) samples. Since the procedure is slightly different for solid and liquid samples, we begin with solids.

Once the sample is correctly aligned, and the temperature is stabilized, we take spectra of the background irradiance with the two spectrometers in order to subtract them through Eq. (1). Then, we take interferograms with the equalization wavelength set in the red band of the short-wavelength spectrometer. Typical interferograms with the background irradiance removed are presented in Fig. 7, showing an oscillatory pattern with a varying frequency. The frequency is larger at the blue end, where dispersion is higher, and decreases as the signal approaches the stationary point. As we move further, the frequency increases again up to NIR wavelengths where it starts decreasing slowly. It is evident that resolving fringes becomes more challenging for short wavelengths. This is why we must use a prism spectrometer which has small dispersion in this region (see Fig. 5).

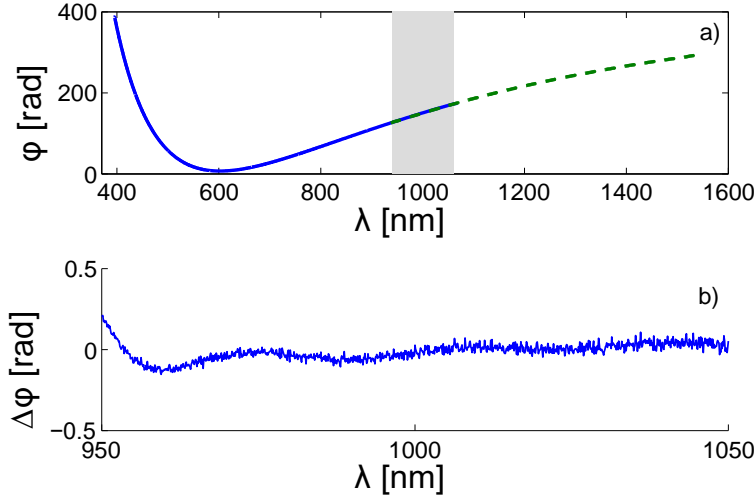

Fig. 8 (a) Phase obtained from the spectrograms shown in Fig. 7 reconstructed using the overlapping zone (shadowed region). (b) Difference of the phase values obtained with the two spectrometers in the overlapping zone.

The next step is to remove the sample and take another spectrogram with the third spectrometer in order to determine the displacement between the interferometer mirrors, *l*. The sample thickness is measured separately with a high-resolution (0.1 μm) micrometer.

In the processing stage, the phases below 1050 nm and above 950 nm are extracted from the short- and long-wavelength spectrograms, respectively, following the procedure described in Section 2. We note that, after performing the unwrapping method, and due to the $2k\pi$ ambiguity, there is a jump between the phases obtained from the two spectrograms. We use the overlapping zone between 950 and 1050 nm to obtain a continuous phase. To this end, we compare phases with different $2\pi$ jumps and minimize the distances between the phases in the overlapping zone. In Fig. 8, we show the result for the spectrograms illustrated in Fig. 7. Also shown is the difference between phases in the overlapping zone after performing the adjustment; the mean is 0.014, and the standard deviation is 0.05. Once this adjustment has been completed, we measure the refractive index for the sodium line in a multi-wavelength Abbe refractometer and use it to obtain the value of *k*. This enables us to correct the phase and retrieve the refractive index in the entire spectral range. To test the validity of the result, we measure the refractive index with the Abbe refractometer at other four lines in the visible range. The observed differences are typically $\sim 10^{-4}$ (see Fig. 9), which is the resolution of our Abbe refractometer. At the end, in order to smooth the data and to describe the dispersion curve in a compact way, we fit the retrieved refractive index to a five-term Cauchy relation:

$$n^2 - 1 = a_0 + a_1\lambda + a_2\lambda^{-2} + a_3\lambda^4 + a_4\lambda^2 + a_2\lambda^4 \ . \tag{5}$$

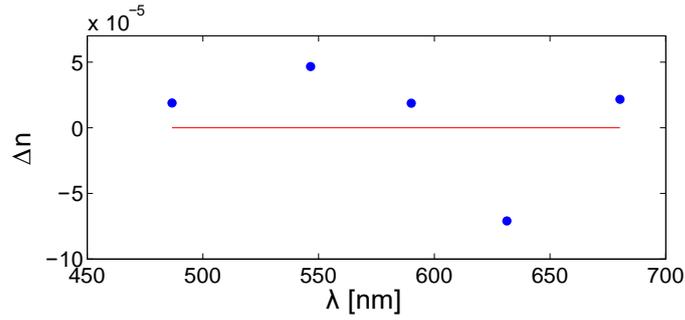

Fig. 9 Deviations of the calculated refractive index from the value measured with an Abbe refractometer. The sample is a BK7 window, and the measurements were taken at wavelengths of 486, 546, 589, 633, and 680 nm.

The measurements of liquid samples follow a slightly different procedure, because the liquid has to be placed inside a cell, and the presence of the cell walls introduces additional terms in Eq. (1b). Our cell is made of quartz and has 1.25 mm thick walls. To partially compensate for this effect, we place a 2 mm quartz sample in the reference arm of the Michelson interferometer and measure the phase difference with the empty cell in order to subtract it from the phase acquired for the cell filled with the liquid sample.

In Table 1, we display the sample thicknesses and the displacement of the interferometer mirrors in each measurement.

Table 1. Sample thickness and displacement between mirrors

| Sample | Thickness* [μm] | Displacement [μm] |
|---|---|---|
| Quartz | 2052.7 ± 0.2 | 979.8 ± 0.3 |
| BK7 | 1029.0 ± 0.2 | 555.4 ± 0.3 |
| Water | 1026.3* ± 0.2 | 360.4 ± 0.3 |
| Ethanol | 1026.3* ± 0.2 | 386.9 ± 0.3 |

* For liquids, the sample thickness corresponds to the inner thickness of the cell.

## 5. Results

In Fig. 10, we show the refractive and group index dispersion curves for different samples; the Cauchy coefficients for the refractive index are displayed in Table 2. The liquid samples were measured at a temperature of 20 °C and ambient pressure, whereas measurements on the solid samples were performed at ambient conditions. The refractive index curves exhibit the typical features of dispersion in transparent materials: high refractive index and large dispersion at short wavelengths and low refractive index and small dispersion in the IR region. Of note is that the group index has a minimum at IR wavelengths; this is associated with the transition between the normal and anomalous dispersion regimes.

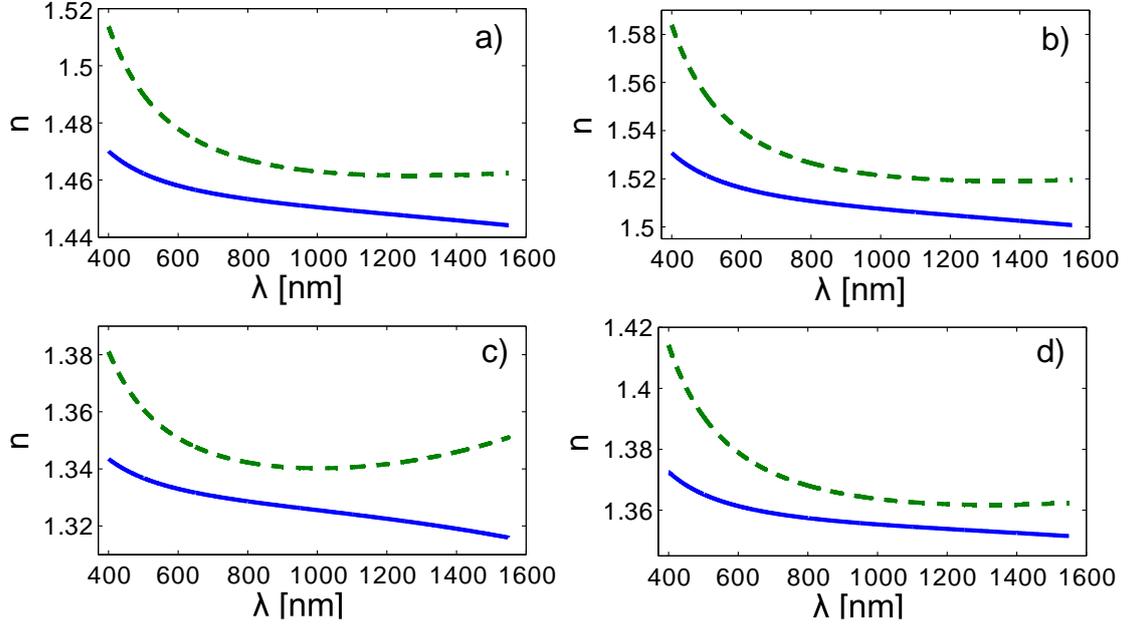

Fig. 10 Refractive (solid line) and group (dashed line) indices of fused silica (a), BK7 (b), water (c), and ethanol (d).

To test the quality of the obtained refractive index curves, we compare our results with refractive index values found in the literature [28]. The deviations are plotted in Fig. 11 and exhibit values of $\sim 10^{-4}$. Furthermore, several measurements were taken for each sample on different days with deviations of $10^{-5}$.

Table 2. Dispersion coefficients (Eq. 5) for the measured refractive indices and their standard deviations ($\sigma$) compared to [28]

|  | $a_0$ | $a_1\,[nm^2]$ | $a_2\,[nm^4]$ | $a_3\,[nm^{-2}]$ | $a_4\,[nm^{-4}]$ | $\sigma$ |
|---|---|---|---|---|---|---|
| Quartz | 1.1044 | $8.621 \cdot 10^3$ | $1.120 \cdot 10^8$ | $-9.120 \cdot 10^{-9}$ | $-8.200 \cdot 10^{-17}$ | $3.78 \cdot 10^{-5}$ |
| BK7 | 1.2718 | $1.060 \cdot 10^4$ | $1.708 \cdot 10^8$ | $-1.023 \cdot 10^{-8}$ | $8.352 \cdot 10^{-17}$ | $8.32 \cdot 10^{-5}$ |
| Water | 0.7608 | $6.788 \cdot 10^3$ | $1.255 \cdot 10^7$ | $-8.561 \cdot 10^{-9}$ | $-1.969 \cdot 10^{-15}$ | $8.51 \cdot 10^{-5}$ |
| Ethanol | 0.8323 | $7.604 \cdot 10^3$ | $1.159 \cdot 10^8$ | $-2.611 \cdot 10^{-9}$ | $-4.518 \cdot 10^{-16}$ | $1.13 \cdot 10^{-4}$ |

We now estimate the precision of the refractive index measurement. The uncertainty in the refractive index is due to several factors including phase measurement, phase ambiguity, spectrometer calibration, path difference, and sample width measurement. In Table 3, we summarize these different contributions. The sum of all contributions is given by

$$\Delta n = \sqrt{\left(\frac{\lambda}{4\pi d}\Delta\varphi_u\right)^2 + \left(\frac{\lambda}{2d}\Delta k\right)^2 + \left[\frac{n-n_g(\lambda_{eq})}{\lambda}\Delta\lambda\right]^2 + \left(\frac{\Delta l}{d}\right)^2 + \left(\frac{n-1}{d}\Delta d\right)^2} . \quad (7)$$

The phase uncertainty is related to errors in the measured irradiance. It is higher at the maxima and minima, where the slope is zero. It has been shown [10] that, at these points, the phase uncertainty is $\Delta\varphi_u = 2/\sqrt{\Delta I}$, where $\Delta I = I_{max} - I_{min}$. In our instrument, $\Delta I$ is smaller at short wavelengths for both phase measurement spectrometers (for example, see Fig. 2). It can be as low as $10^3$, even with a 16-bit camera, resulting in a phase error of 0.06 rad. This gives the refractive index uncertainty of $7 \times 10^{-5}$ at the wavelength $\lambda$ = 1550 nm and sample thickness of 1 mm.

The uncertainty in the value of *k* is related to the precision of the Abbe refractometer. Assuming it is $10^{-4}$, from the measurement at the sodium line ($\lambda$ = 589 nm), and d = 1 mm as before, we obtain $\Delta k = \text{nint}(2d\Delta n/\lambda) = 0$. However, $\Delta k = 1$ for d $\cong$ 1.5 mm, which gives $\Delta n$ = 5 × $10^{-4}$ for a limiting wavelength of 1550 nm. From Fig. 11, we see that such deviations were never obtained in our measurement. Therefore, we conclude that phase ambiguity does not influence the refractive index uncertainty in our measurement.

Regarding wavelength calibration, the use of a spectral lamp combined with SRWLI enables us to obtain the relative uncertainty $\Delta\lambda/\lambda$ below $3 \times 10^{-4}$. Considering the difference of the refractive index and group index of less than 0.05, the contribution to the refractive index uncertainty is below $1.5 \times 10^{-5}$.

The next contribution to the refractive index uncertainty is the path difference in air between the two arms of the interferometer. Our measurement with SRWLI gives an uncertainty of 0.3 μm, which contributes to $\Delta n = 3 \times 10^{-4}$ for a sample thickness of 1 mm.

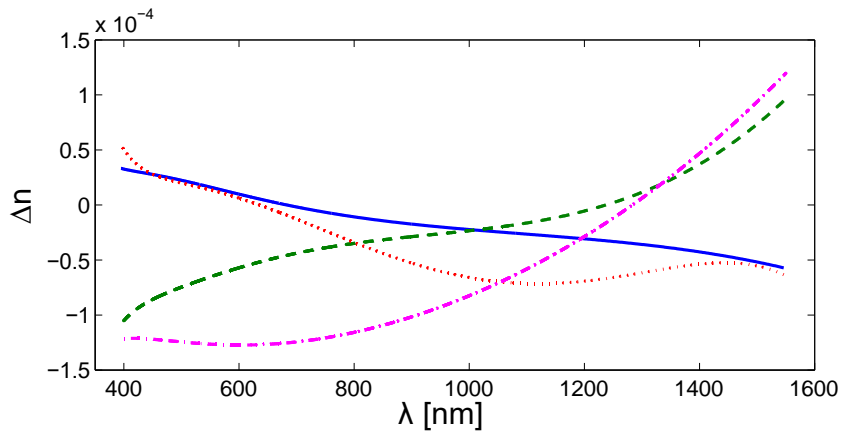

Fig. 11 Deviation of the refractive indices of quartz (solid line), BK7 (dashed line), water (dotted line), and ethanol (dashed-dotted line) from the reference values reported in [28].

The last contribution is given by the measurement of the sample thickness. We measure the thickness of solid samples with a high-precision micrometer with 0.1 µm resolution. However, the accuracy of the instrument is 0.2 µm. Taking d = 1 mm and n = 1.5, we obtain $\Delta n = 10^{-4}$, while for a higher refractive index of n = 2, $\Delta n = 2 \times 10^{-4}$. For liquid samples, we must measure the inner and outer thickness of the sample length. The outer thickness is measured with the same micrometer, while the inner thickness is measured with SRWLI [29]. This last measurement increases the contribution to the refractive index uncertainty to $2 \times 10^{-4}$ for n = 1.5.

In brief, main contributions to the refractive index uncertainty come from measurement of the path length difference in the interferometer and sample thickness. These two contributions decrease for thicker samples; however, thicker samples can increase the weight of the phase ambiguity uncertainty.

Table 3. Summary of uncertainties for d = 1 mm, n = 1.5, and $\lambda$ = 1550 nm

| Cause | Value | Contribution to n |
| --- | --- | --- |
| Phase measurement | 0.06 | $7 \times 10^{-5}$ |
| Phase ambiguity | 0 | 0 |
| Wavelength calibration | $3 \times 10^{-4}$* | $1.5 \times 10^{-5}$ |
| Mirror displacement | 0.3 µm | $3 \times 10^{-4}$ |
| Sample thickness | 0.1–0.2 µm** | $1 \times 10^{-4} - 2 \times 10^{-4}$** |

\* Relative value
\*\* For solids and liquids, respectively

**6. Conclusions**

SRWLI was proved to be a successful technique to retrieve refractive and group indices in a wide frequency range of 400–1550 nm, i.e. nearly two octaves, with several spectrograms. Crucial to this achievement is the use of a prism spectrometer in the VIS region that enables us to resolve fringes in the high-dispersion blue spectral region. Furthermore, SRWLI was applied to calibration of the spectrometers and measurement of the displacement in air between the mirrors in the Michelson interferometer. Comparison with data available in literature shows that the results for solid and liquid samples deviate by $10^{-4}$ or less in the entire spectral range, while the standard uncertainty analysis shows that major uncertainty contributions come from the distance measurements, sample thickness, and mirror displacement, with values of 0.03%. Our instrument includes a heat exchanger that not only allows controlling the temperature of the samples but also permits measurement of the dispersion as a function of temperature.

In the future, we plan to complement our instrument with new devices that would enable us to take measurements in the UV spectral range. Additionally, we plan to measure absorption data with high precision using the same instrument.

**Funding**

Ministerio de Economía y Competitividad (MINECO) (MAT2014-57943-C3-2-P); Xunta de Galicia and FEDER (AGRU 2015/11 and GRC ED431C 2016/001).